# DIFFUSION REGIME OF ELECTRON-ELECTRON COLLISIONS IN WEAKLY IONIZED PLASMAS

B. Breizman[1], G. Stupakov[2] and G. Vekstein[3,a]

[1]Institute for Fusion Studies, The University of Texas, Austin, Texas 78712, USA

[2]SLAC National Accelerator Laboratory, Menlo Park, California 94025, USA

[3]Jodrell Bank Centre of Astrophysics, The University of Manchester, Manchester M13 9PL, UK

[a]Electronic mail: g.vekstein@manchester.ac.uk

## ABSTRACT

We consider weakly ionized plasma where frequent elastic scattering of electrons on neutrals change the individual acts and the rate of electron-electron collisions significantly. In this case, the kinetics of electron thermalization is very different from that in fully ionized plasma. The colliding electrons do not move freely. They, instead, diffuse because of fast scattering on neutrals. We demonstrate how a proper account of this diffusion enables one to estimate the characteristic time of electron thermalization. We also present a rigorous derivation of the kinetic equation for electrons by using Bogolyubov's method based on Liouville equations for multi-particle distribution functions.

## 1. INTRODUCTION

In a recent publication [1], in which the authors discussed the impact of electron-electron collisions on the conductivity of weakly ionized plasmas, they pointed out that frequent scattering of electrons on neutral particles can modify the interaction between the electrons themselves. The mode of electron-electron collisions changes when the mean-free path for electron-neutral scattering, $\lambda_{en}$, is shorter than the distance between the colliding electrons. For thermal electrons separated by a distance of order of the Debye radius $r_D$, the corresponding electron-neutral collision frequency $\nu_{en}$ should then exceed the electron plasma frequency $\omega_p = \sqrt{4\pi n e^2 / m}$, where $n$ and $m$ are the electron number density and mass. The situation with

$$\lambda_{en} \ll r_D , \qquad (1)$$

or, equivalently,

$$\nu_{en} \gg \omega_p \qquad (2)$$

is quite common in low-temperature plasmas [1]. In this case, frequent electron-neutral collisions limit electron mobility. However, as demonstrated numerically in Ref. [1], rarer electron-electron collisions still have a noticeable impact on plasma conductivity. Although electron-electron collisions do not change the net momentum of electrons, they do affect electron mobility via modification of the electron energy distribution. This effect is well-known in fully ionized plasma [2], and it takes place in partially ionized plasmas as well.

We note that the electron-neutral collisions are primarily elastic in low-temperature plasmas, and they do not change electron energies to the lowest order in mass ratio. That is why electron-electron collisions govern the electron energy distribution despite a relatively low electron density. From the theory perspective, relaxation of the electron energy distribution via electron-electron collisions in weakly ionized plasma deserves attention as a stand-alone problem that we address in this paper. We first consider a qualitative picture of the process and the resulting estimate for the relaxation timescale. We then present a rigorous derivation of the kinetic equation for the electron distribution function. This reveals to what extent frequent scattering of electrons by neutrals hinder the energy exchange between the colliding electrons.

In Ref. [1] this issue has been discussed qualitatively in terms of conventional characteristics of binary electron-electron collisions such as the impact parameter, the scattering angle, and the cross section. However, their relevance to our case of interest looks questionable. Indeed, due to frequent scattering on neutrals, the electrons have diffusion-type trajectories sketched in Fig.1 as computer simulated random walks.

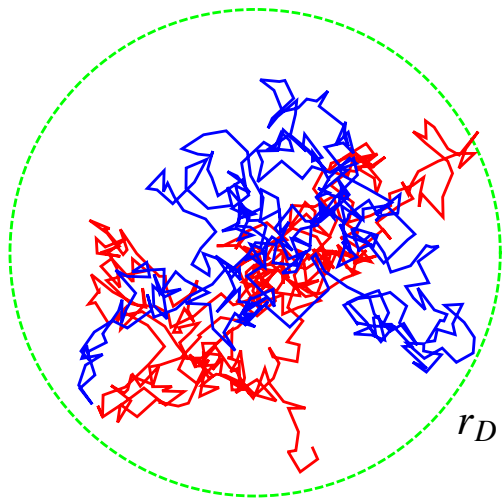

Fig. 1. Schematic view of electron-electron collision affected by frequent scattering on neutrals. Shown are two trajectories (blue and red) of the electrons colliding within the Debye sphere (green dashed line) for the case of $\sqrt{\langle \lambda_{en}^2 \rangle} = 0.08 r_D$.

Consequently, the relative velocity of any two electrons changes direction and magnitude many times in the course of their Coulomb interaction. As a result, the very notions of the collision impact parameter, the scattering angle, and the scattering cross-section become unclear. Instead, the diffusive motion of electrons needs to be recognized explicitly in this situation.

## 2. QUALITATIVE CONSIDERATION

Consider two electrons separated by some distance $r$. Due to spatial diffusion, their separation increases with time, and it takes roughly the time $(\Delta t)_{ee} \sim r^2 / D_e$ to double it, where $D_e$ is the diffusion coefficient. Therefore, $(\Delta t)_{ee}$ is the effective interaction time of these electrons. Then, since $D_e \sim u_e \lambda_{en} \sim u_e^2 \nu_{en}^{-1}$, where $u_e$ is the characteristic velocity, the interaction time can be written as

$$(\Delta t)_{ee} \sim r^2 \nu_{en} / u_e^2 \, . \tag{3}$$

Furthermore, during this time electrons experience about

$$N_{en} \sim \nu_{en} (\Delta t)_{ee} \sim r^2 \nu_{en}^2 / u_e^2 \gg 1 \tag{4}$$

collisions with neutrals, which divides the total interaction time $(\Delta t)_{ee}$ into $N_{en}$ sub-intervals of an un-interrupted motion under the Coulomb force. During a single sub-interval that lasts $(\Delta t)_{en} \sim \nu_{en}^{-1}$ the electron acquires a velocity increment

$$\delta \mathbf{u}_e \sim \frac{e^2}{mr^2 \nu_{en}} \frac{\mathbf{r}}{r}$$

so that the electron kinetic energy changes by

$$\delta E = m \mathbf{u}_e \cdot \delta \mathbf{u}_e \sim \frac{e^2}{r^3 \nu_{en}} (\mathbf{u}_e \cdot \mathbf{r}) = \frac{e^2 u_e}{r^2 \nu_{en}} \cos\theta ,$$

where $\theta$ is the angle between the vectors $\vec{r}$ and $\vec{u}_e$. This angle changes randomly from one sub-interval to another. Hence, the electron undergoes a random walk along the energy axis with a step of order of

$$\delta E \sim e^2 u_e / r^2 \nu_{en} \, . \tag{5}$$

The amount of energy exchanged in the course of a single electron-electron collision can, therefore, be estimated as

$$\Delta E \sim \delta E \cdot \sqrt{N_{en}} \sim e^2 / r \tag{6}$$

Although this estimate follows immediately from the energy conservation (it is apparently the characteristic potential energy of the interacting electrons), its derivation here emphasizes the diffusive aspect of the electron-electron interaction.

The distance $r$ between two interacting electrons is bounded from above by the Debye radius $r_D$, and from below by the distance of closest approach $r_0 \sim e^2 / m u_e^2$. For thermal electrons $r_0 \sim r_D / N_D$, where $N_D = nr_D^3$ is the Debye number. We assume that the plasma is nearly ideal, i.e., $N_D \gg 1$.

The consequences of diffusion depend on the ratio of the electron-neutral mean-free path $\lambda_{en}$ to the Debye radius. This ratio separates three limiting cases. In the first case, the mean-free path $\lambda_{en}$ is longer than the Debye radius. In this limit, the electron-electron collisions are not affected by neutrals at all, and the electron energy relaxation time $\tau_1$ can be estimated as the one in the fully ionized plasma [3]

$$\tau_1 \sim \frac{N_D}{\omega_p \Lambda}, \tag{7}$$

where $\Lambda = \ln(r_D / r_0) = \ln N_D$ is the standard Coulomb logarithm.

The second, intermediate, case takes place when $\lambda_{en}$ is shorter than $r_D$ but still long compared with $r_0 = r_D / N_D$. In terms of the elastic scattering frequency $\nu_{en}$ it reads

$$\omega_p < \nu_{en} < N_D \omega_p . \tag{8}$$

In this case, two kinds of electron-electron collisions take place. If the distance $r$ between the electrons is within the interval $\lambda_{en} < r < r_D$, the electron trajectories are diffusive. On the other hand, when $r_0 < r < \lambda_{en}$, the role of diffusion is negligible. The time-scale $\tau_2$ of the energy equilibration due to the latter type of collisions follows directly from Eq. (7), where the classical Coulomb logarithm $\Lambda$ should be replaced by the modified one [1], $\tilde{\Lambda}$, as $\tilde{\Lambda} = \ln(\lambda_{en} / r_0) = \ln(N_d \omega_p / \nu_{en})$. Thus,

$$\tau_2 \sim \frac{N_D}{\omega_p \tilde{\Lambda}} \sim \frac{N_D}{\omega_p \ln(N_D \omega_p / \nu_{en})} . \tag{9}$$

With regard to the diffusion-type collisions, we note that the energy acquisition (6) (which for an individual electron could be either positive or negative) is small compared to the electron kinetic energy in the case of $r > r_0$. The kinetic energies of the involved electrons will follow the diffusion law. The collisions within the interval $(r, r+dr)$ provide the following contribution to the energy diffusion coefficient $D_E$: $dD_E = \frac{[\Delta E(r)]^2}{(\Delta t)_{ee}} 4\pi r^2 n dr$, with $\Delta E(r)$ and $\Delta t_{ee}(r)$ given by equations (6) and (3) respectively. We thus obtain

$$D_E = \int_{\lambda_{en}}^{r_D} dD_E = \frac{\omega_p^2 m u_e^2 e^2}{\nu_{en}} \int_{\lambda_{en}}^{r_D} \frac{dr}{r^2} = (mu_e^2)^2 \frac{\omega_p}{N_D} \tag{10}$$

This diffusion coefficient is determined by collisions with $r \sim \lambda_{en}$ and, therefore, it does not depend on the electron scattering frequency $\nu_{en}$. The corresponding time-scale of energy exchange is

$$\tau_2^{(diff)} \sim \frac{(mu_e^2)^2}{D_E} \sim \frac{N_D}{\omega_p} \quad (11)$$

As seen from Eqs. (11) and (9), the energy equilibration of electrons is predominantly due to classical electron-electron collisions as long as $\lambda_{en} > r_0$, and, hence, the modified Coulomb logarithm $\tilde{\Lambda}$ in (9) is large.

However, further increase of the electron scattering frequency shrinks the domain of classical collisions. This happens when $\nu_{en} > N_D \omega_p$ and $\lambda_{en}$ is shorter than $r_0$. All electron-electron collisions are in the diffusion regime in this case (case 3).

The major role is played then by collisions with the shortest separation, which in this case is $r \sim r_0$. However, unlike the case 2, the amount of energy exchanged in the course of the electron-electron collision with $r \sim r_0$ is comparable to the total kinetic energy of electrons [see Eq.(6)( Note, though, that such a "single" collision comprises a large number [see Eq.(4)] of mini-shots interrupted by the electron scattering on neutrals. Thus, the electron energy equilibration time in this case, $\tau_3$, can be estimated as $\tau_3 \sim \Delta t_{ee}(r_0)/N_0$, where $N_0 \sim n r_0^3 \sim N_D^{-2}$ is the number of electrons inside the sphere of radius $r_0$, and $\Delta t_{ee}(r_0)$ is given by Eq.(3). We finally obtain

$$\tau_3 \sim \frac{\nu_{en}}{\omega_p^2} \,. \quad (12)$$

As seen from Eq. (12), in the case of very frequent scattering of electrons on neutrals the dependence of $\tau_3$ on the energy of electrons is determined entirely by how the scattering frequency varies with the energy of electrons. It is well-known (see, e.g., Ref. [4]) that the cross–section of this scattering is typically almost constant within the energy range relevant to weakly ionized plasmas. Then, according to Eq. (12), the energy equilibration time of electrons is increases with temperature as $\tau_3 \propto T_e^{1/2}$, but much slower than in the case of fully ionized plasma where $\tau_1 \propto T_e^{3/2}$.

To conclude this section, we recall that we have so far neglected the energy exchange in the process of electron scattering by neutral. This is justified if the respective energy exchange time, $\tau_{en}^{(E)} \sim \nu_{en}^{-1} \frac{M_n}{m}$, where $M_n$ is the mass of neutrals, exceeds the electron-electron energy equilibration time (12), i.e. $\nu_{en} < \omega_p (M_n/m)^{1/2}$. On the other hand, the very realization of the diffusion regime of electron-electron collisions requires $\nu_{en} > \omega_p N_D$. These two constraints are compatible if $N_D < (M_n/m)^{1/2}$, which is not a very restrictive condition, because a typical Debye number is moderate ($N_D \sim 10^2$) in low-temperature plasmas such as discussed, for example, in Ref. [1].

## 3. KINETIC EQUATION

As shown in Section 2, the role of elastic scattering is most pronounced in the limit of $\lambda_{en} \ll r_0$ when all collisions are in the diffusive regime. We herein present a derivation of the kinetic equation for this limiting case.

Our starting point is a kinetic equation for the two-particle distribution function (see, e.g., Ref. [5]) of the interacting electrons, $f^{(2)}(t;\mathbf{r}_1;\mathbf{r}_2;\mathbf{p}_1;\mathbf{p}_2)$. We take into account that the plasma is nearly ideal ($N_D \gg 1$) and neglect a contribution of the three-particle function $f^{(3)}$. We, however, include elastic collisions of electrons with the background neutrals. Such collisions are governed by the elastic scattering operators so that the kinetic equation for $f^{(2)}$ takes the form:

$$\frac{\partial f^{(2)}}{\partial t} + \frac{\mathbf{p}_1}{m}\frac{\partial f}{\partial \mathbf{r}_1} + \frac{\mathbf{p}_2}{m}\frac{\partial f^{(2)}}{\partial \mathbf{r}_2} - \frac{\partial U}{\partial \mathbf{r}_1}\frac{\partial f^{(2)}}{\partial \mathbf{p}_1} - \frac{\partial U}{\partial \mathbf{r}_2}\frac{\partial f^{(2)}}{\partial \mathbf{p}_2} =$$
$$-\int n_0 \frac{p_1}{m}\frac{d\sigma}{do'_1}\Big[f^{(2)}(t;p_1;p_2;\mathbf{n}_1;\mathbf{n}_2) - f^{(2)}(t;p_1;p_2;\mathbf{n}'_1;\mathbf{n}_2)\Big]do'_1 \qquad (13)$$
$$-\int n_0 \frac{p_2}{m}\frac{d\sigma}{do'_2}\Big[f^{(2)}(t;p_1;p_2;\mathbf{n}_1;\mathbf{n}_2) - f^{(2)}(t;p_1;p_2;\mathbf{n}_1;\mathbf{n}'_2)\Big]do'_2$$

Here $U \equiv e^2/|\mathbf{r}_1 - \mathbf{r}_2|$ is the Coulomb potential, $n_0$ is the number density of the background neutrals, $d\sigma/do$ is the differential cross-section for the elastic scattering, $do$ is the solid angle element in the momentum space, and $\mathbf{n}_1 \equiv \mathbf{p}_1/p_1$ and $\mathbf{n}_2 \equiv \mathbf{p}_2/p_2$ are the unit vectors. We note that the arguments of $d\sigma/do'_1$ are $p_1$ and $(\mathbf{n}_1 \cdot \mathbf{n}'_1)$, whereas the arguments of $d\sigma/do'_2$ are $p_2$ and $(\mathbf{n}_2 \cdot \mathbf{n}'_2)$.

Frequent elastic collisions make the distribution function nearly isotropic in momentum space. We, therefore, consider $f^{(2)}$ as having a large isotropic part $F(t;\mathbf{r}_1;\mathbf{r}_2;p_1;p_2)$ and a small anisotropic correction $g(t;\mathbf{r}_1;\mathbf{r}_2;p_1;p_2;\mathbf{n}_1;\mathbf{n}_2)$, i.e.,

$$f^{(2)} \approx F(t;\mathbf{r}_1;\mathbf{r}_2;p_1;p_2) + g(t;\mathbf{r}_1;\mathbf{r}_2;p_1;p_2;\mathbf{n}_1;\mathbf{n}_2) \qquad (14)$$

The anisotropic correction is implied to have a zero angular average value, i.e.

$$\langle g(t;\mathbf{r}_1;\mathbf{r}_2;p_1;p_2;\mathbf{n}_1;\mathbf{n}_2)\rangle = 0 , \tag{15}$$

where angular brackets denote averaging over all directions of $\mathbf{n}_1$ and $\mathbf{n}_2$. We next neglect the small time derivative of $g$ and split Eq. (13) into the following two coupled equations for $F$ and $g$:

$$\frac{\partial F}{\partial t} + \left\langle \frac{\mathbf{p}_1}{m}\frac{\partial g}{\partial \mathbf{r}_1}\right\rangle + \left\langle \frac{\mathbf{p}_2}{m}\frac{\partial g}{\partial \mathbf{r}_2}\right\rangle - \left\langle \frac{\partial U}{\partial \mathbf{r}_1}\frac{\partial g}{\partial \mathbf{p}_1}\right\rangle - \left\langle \frac{\partial U}{\partial \mathbf{r}_2}\frac{\partial g}{\partial \mathbf{p}_2}\right\rangle = 0 \tag{16}$$

$$\frac{\mathbf{p}_1}{m}\frac{\partial F}{\partial \mathbf{r}_1} + \frac{\mathbf{p}_2}{m}\frac{\partial F}{\partial \mathbf{r}_2} - \frac{\partial U}{\partial \mathbf{r}_1}\frac{\partial F}{\partial \mathbf{p}_1} - \frac{\partial U}{\partial \mathbf{r}_2}\frac{\partial F}{\partial \mathbf{p}_2} =$$
$$-\int n_0 \frac{p_1}{m}\frac{d\sigma}{do'_1}\left[g(t;p_1;p_2;\mathbf{n}_1;\mathbf{n}_2) - g(t;p_1;p_2;\mathbf{n}'_1;\mathbf{n}_2)\right]do'_1 \tag{17}$$
$$-\int n_0 \frac{p_2}{m}\frac{d\sigma}{do'_2}\left[g(t;p_1;p_2;\mathbf{n}_1;\mathbf{n}_2) - g(t;p_1;p_2;\mathbf{n}_1;\mathbf{n}'_2)\right]do'_2$$

Equation (17) has a straightforward solution of the form

$$g = -\lambda(p_1)\left[\left(\mathbf{n}_1\frac{\partial F}{\partial \mathbf{r}_1}\right) - \left(\frac{\partial U}{\partial \mathbf{r}_1}\mathbf{n}_1\right)\frac{m}{p_1}\frac{\partial F}{\partial p_1}\right] - \lambda(p_2)\left[\left(\mathbf{n}_2\frac{\partial F}{\partial \mathbf{r}_2}\right) - \left(\frac{\partial U}{\partial \mathbf{r}_2}\mathbf{n}_2\right)\frac{m}{p_2}\frac{\partial F}{\partial p_2}\right], \tag{18}$$

where $\lambda(p) = \left[\int n_0 \frac{d\sigma}{do'}(1 - \mathbf{n}\cdot\mathbf{n}')\,do'\right]^{-1}$ is the electron mean free path for elastic scattering on neutrals. In what follows, we put $\lambda_1 \equiv \lambda(p_1)$, $\lambda_2 \equiv \lambda(p_2)$ for brevity.

Substitution of $g$ into Eq. (16) reduces this equation to

$$\frac{\partial F}{\partial t} - \frac{1}{3}\frac{\partial}{\partial x_{1i}}\lambda_1\left[\frac{p_1}{m}\frac{\partial F}{\partial x_{1i}} - \frac{\partial U}{\partial x_{1i}}\frac{\partial F}{\partial p_1}\right] - \frac{1}{3}\frac{\partial}{\partial x_{2i}}\lambda_2\left[\frac{p_2}{m}\frac{\partial F}{\partial x_{2i}} - \frac{\partial U}{\partial x_{2i}}\frac{\partial F}{\partial p_2}\right]$$
$$+ \frac{1}{3p_1^2}\frac{\partial}{\partial p_1}\frac{\partial U}{\partial x_{1i}}mp_1\lambda_1\left[\frac{p_1}{m}\frac{\partial F}{\partial x_{1i}} - \frac{\partial U}{\partial x_{1i}}\frac{\partial F}{\partial p_1}\right] + \frac{1}{3p_2^2}\frac{\partial}{\partial p_2}\frac{\partial U}{\partial x_{2i}}mp_2\lambda_2\left[\frac{p_2}{m}\frac{\partial F}{\partial x_{2i}} - \frac{\partial U}{\partial x_{2i}}\frac{\partial F}{\partial p_2}\right] = 0 \tag{19}$$

In a spatially uniform and isotropic background, the distribution function $F$ depends only on distance $r \equiv |\mathbf{r}_1 - \mathbf{r}_2|$ between the particles rather than on $\mathbf{r}_1$ and $\mathbf{r}_2$ separately. Equivalently, one can consider $F$ as being a function of $U$ and transform Eq. (19) to

$$\frac{\partial F}{\partial t} = \frac{U^4}{3e^4}\frac{\partial}{\partial U}\lambda_1\left[\frac{p_1}{m}\frac{\partial F}{\partial U} - \frac{\partial F}{\partial p_1}\right] + \frac{U^4}{3e^4}\frac{\partial}{\partial U}\lambda_2\left[\frac{p_2}{m}\frac{\partial F}{\partial U} - \frac{\partial F}{\partial p_2}\right]$$
$$-\frac{U^4}{3e^4 p_1^2}\frac{\partial}{\partial p_1}mp_1\lambda_1\left[\frac{p_1}{m}\frac{\partial F}{\partial U} - \frac{\partial F}{\partial p_1}\right] - \frac{U^4}{3e^4 p_2^2}\frac{\partial}{\partial p_2}mp_2\lambda_2\left[\frac{p_2}{m}\frac{\partial F}{\partial U} - \frac{\partial F}{\partial p_2}\right] \quad (20)$$

It follows from Eq. (20) that $\frac{\partial F}{\partial t} = 0$ for any function $F(E)$, where $E = \frac{p_1^2}{2m} + \frac{p_2^2}{2m} + U(r)$ is the conserved total energy of the two interacting particles. However, our interest is only in the two-particle function $F(p_1, p_2, r)$ that reduces to the product of two single-particle distribution functions $f(p)$ as $F(p_1, p_2, r) = f(p_1) \cdot f(p_2)$ where $U(r) \ll E$. The only stationary function of this kind is $F(E) \propto \exp(-E/T)$, which corresponds to thermodynamic equilibrium.

Furthermore, we note that the first two terms on the right hand side of Eq.(20) cancel each other if $F$ is a function of only two variables: the energy $E$ and a new variable $q$ defined as

$$q(p_1, p_2) \equiv \int_{p_2}^{p_1} \frac{dp}{\lambda(p)} \quad . \quad (21)$$

It is, therefore, convenient to re-wright Eq. (20) in terms of $E$, $q$ and $U$, which reduces this equation to the following compact form:

$$\frac{\partial F}{\partial t} = \frac{(p_1\lambda_1 + p_2\lambda_2)}{3mp_1^2 p_2^2 \lambda_1 \lambda_2}\frac{U^4}{e^4}\left[\frac{\partial}{\partial U}\lambda_1\lambda_2 p_1^2 p_2^2 \frac{\partial F}{\partial U} + m^2\frac{\partial}{\partial q}(p_1 p_2)\frac{\partial F}{\partial q}\right], \quad (22)$$

where $p_1$ and $p_2$ are now functions of $E$, $q$ and $U$.

In what follows, we assume that the elastic mean free path of electrons does not depend on their energy, i.e., $\lambda_1 = \lambda_2 \equiv \lambda_0$ (this is actually the case at low energies [4]). Then, the variable $q$ takes the form $q = (p_1 - p_2)/\lambda_0 \equiv \rho/\lambda_0$, and Eq.(22) reduces to

$$\frac{\partial F}{\partial t} = \frac{m\lambda_0(p_1 + p_2)}{3(p_1 p_2)^2}\frac{U^4}{e^4}\left[\frac{1}{m^2}\frac{\partial}{\partial U}(p_1 p_2)^2\frac{\partial F}{\partial U} + \frac{\partial}{\partial \rho}(p_1 p_2)\frac{\partial F}{\partial \rho}\right]. \quad (23)$$

In terms of dimensionless variables

$$y \equiv \frac{\rho}{\sqrt{2mE}} \,, \quad z \equiv \frac{U}{E}, \quad \tau \equiv t\frac{\omega_p^2 \lambda_0}{\sqrt{2E/m}} \quad (24)$$

we have

$$p_1 + p_2 = \sqrt{2mE}\sqrt{2 - y^2 - 2z}$$
$$p_1 p_2 = mE(1 - z - y^2) \quad , \quad (25)$$

which transforms Eq. (23) to

$$\frac{\partial F}{\partial \tau} = Dz^4 \frac{\sqrt{2-2z-y^2}}{(1-z-y^2)^2} \left[ \frac{\partial}{\partial z}(1-z-y^2)^2 \frac{\partial F}{\partial z} + \frac{1}{2}\frac{\partial}{\partial y}(1-z-y^2)\frac{\partial F}{\partial y} \right]. \quad (26)$$

This is a diffusion-type equation with a parameter $D = \frac{2E^3}{3me^4\omega_p^2}$ being very large. Indeed, at $E \sim T$, the thermal energy of electrons, we have

$$D \sim N_D^2 \gg 1 \quad (27)$$

Such a strong diffusion is sufficient to make the function $F(E,y,z)$ nearly independent of $z$ within the relaxation time span of $\Delta\tau \sim 1$. The relevant interval of $z$ is $z_1 < z < 1-y^2$, where $z_1$ can be roughly estimated by equating the time derivative term to the first term on the right hand side of Eq. (26), which gives

$$z_1 \sim D^{-1/2} \sim N_D^{-1} \quad (28)$$

In terms of the distance $r$ this interval corresponds to $r_0 \leq r \leq r_1 \sim r_0 N_D$. On the other hand, in a uniform plasma under consideration, it is sufficient to consider the function $F$ only inside the per-particle volume $\Delta V = 1/n = \frac{4\pi}{3}<r>^3$. This sub-volume defines the maximal separation distance $r_{max} = <r> \sim r_0 N_D^{2/3}$, which is much shorter than $r_1$ and translates into $z_{min} = U(<r>)/E = e^2(4\pi n/3)^{1/3}/E \sim N_D^{-2/3} \gg z_1$. It is, therefore, allowable to consider $F$ as a $z$-independent function of $E$ and $y$ to lowest order.

It should be noted that while the second diffusive term on the right hand side of Eq.(26) plays no role at $z \ll 1$, it is significant at $z \sim 1$. Therefore, the dependence of $F$ on $y$ should be very weak at $z \sim 1$ to match the estimated relaxation time-scale (12).

This feature of the function $F$ can be explored in the following way. It follows from Eq.(26) that

$$\frac{1}{D}\int_{z_{min}}^{(1-y^2)} \frac{\partial F}{\partial \tau}\frac{(1-z-y^2)^2}{z^4\sqrt{2-z-y^2}}dz = \int_{z_{min}}^{(1-y^2)} dz\frac{\partial}{\partial z}(1-z-y^2)^2\frac{\partial F}{\partial z} + \frac{1}{2}\int_{z_{min}}^{(1-y^2)} dz\frac{\partial}{\partial y}(1-z-y^2)\frac{\partial F}{\partial y} \quad (29)$$

The first integral on the right hand side of Eq. (29) vanishes, while the other two integrals, where $F$ may be assumed as $z$-independent, yield the following evolution equation

$$\frac{1}{3Dz_{min}^3}\frac{\partial F}{\partial \tau}=\frac{3}{2}\frac{\partial F}{\partial \tau}=\frac{\sqrt{2-y^2}}{4(1-y^2)^2}\frac{\partial}{\partial y}(1-y^2)^2\frac{\partial F}{\partial y} \ . \qquad (30)$$

We now return to dimensional variables and apply Eq. (30) to statistically independent particles within most of the per-particle volume where the potential energy is negligibly small and

$$F(p_1,p_2,t)= f(p_1,t)\cdot f(p_2,t) \ . \qquad (31)$$

We note from Eq. (25) that

$$\sqrt{2-y^2}=(p_1+p_2)/\sqrt{2mE}, \ (1-y^2)= p_1 p_2 /mE,$$

and that

$$\frac{\partial}{\partial y}=\frac{\sqrt{2mE}\,p_1 p_2}{(p_1+p_2)}\left(\frac{1}{p_1}\frac{\partial}{\partial p_1}-\frac{1}{p_2}\frac{\partial}{\partial p_2}\right),$$

which transforms Eq.(30) to

$$\frac{\partial F}{\partial t}=\frac{\omega_p^2 m\lambda_0}{6 p_1 p_2}\left(\frac{1}{p_1}\frac{\partial}{\partial p_1}-\frac{1}{p_2}\frac{\partial}{\partial p_2}\right)\frac{p_1^2 p_2^2}{(p_1+p_2)}\left(p_2\frac{\partial F}{\partial p_1}-p_1\frac{\partial F}{\partial p_2}\right) \qquad (32)$$

We finally substitute Eq. (31) into Eq. (32) and integrate Eq. (32) over the momentum space of $p_2$ as $\int d^3 p_2 = 4\pi\int_0^\infty p_2^2 dp_2$. The resulting kinetic equation reads

$$\frac{\partial f(p_1,t)}{\partial t}=\frac{2\pi\omega_p^2 m\lambda_0}{3}\frac{1}{p_1^2}\frac{\partial}{\partial p_1}\int_0^\infty p_2\,dp_2\frac{p_1^2 p_2^2}{(p_1+p_2)}\left[p_2 f(p_2)\frac{\partial f(p_1)}{\partial p_1}-p_1 f(p_1)\frac{\partial f(p_2)}{\partial p_2}\right], \qquad (33)$$

where the normalization $4\pi\int_0^\infty f(p,t)p^2 dp =1$ is imposed.

Equation (33) conserves the number of particles, the total energy, and describes an irreversible relaxation of any initial distribution function to the Maxwellian one. These conservation laws as well as the H-theorem can be proven similarly to those for the Boltzmann kinetic equation [5]. The maxwellization of the electron energy distribution has a characteristic time-scale of

$$\tau_M \sim \frac{p}{m\lambda_0 \omega_p^2}\sim \frac{u/\lambda_0}{\omega_p^2}\sim \frac{\nu_{en}}{\omega_p^2} \ , \qquad (34)$$

which is consistent with the qualitative estimate (12).

ACKNOWLEDGMENTS

The authors are grateful to D. Ryutov for helpful comments.

This work was supported by the U.S. Department of Energy Contract Nos. DEFG02–04ER54742 and DESC0016283.